\documentclass[aps,prl,twocolumn,showpacs,10pt,superscriptaddress,preprintnumbers]{revtex4-1}
\usepackage{amsmath}
\usepackage{graphicx}
\usepackage{amssymb}
\usepackage[colorlinks,citecolor=red]{hyperref}

\usepackage{color}    
\usepackage{ulem}

\newcommand{\beq}{\begin{equation}}
\newcommand{\eeq}{\end{equation}}
\newcommand{\bea}{\begin{eqnarray}}
\newcommand{\eea}{\end{eqnarray}}
\newcommand{\nn}{\nonumber}

\begin{document}

\preprint{
{\vbox {
\hbox{\bf LA-UR-21-20373}
\hbox{\bf MSUHEP-21-001}
}}}
\vspace*{0.2cm}

\title{The anomalous $Zb\bar{b}$ couplings: From LEP to LHC}

\author{Bin Yan}
\email{binyan@lanl.gov}
\affiliation{Theoretical Division, Group T-2, MS B283, Los Alamos National Laboratory, P.O. Box 1663, Los Alamos, NM 87545, USA}

\author{C.-P. Yuan}
\email{yuan@pa.msu.edu}
\affiliation{Department of Physics and Astronomy,
Michigan State University, East Lansing, MI 48824, USA}

\begin{abstract}
The bottom quark forward-backward asymmetry ($A_{FB}^b$) data at LEP exhibits a  long-standing discrepancy with the standard model prediction. 
We propose a novel method to probe the $Zb\bar{b}$ interactions through $gg\to Zh$ production at the LHC, which is sensitive to the axial-vector component of the $Zb\bar{b}$ couplings. The apparent degeneracy of the anomalous $Zb\bar{b}$ couplings implied by the LEP precision electroweak measurements seems to be resolved by the current 13 TeV LHC $Zh$ data, which is however dominated by the two data points with high transverse momentum of $Z$ boson whose central values are in conflict with the standard model prediction. 
We also show the potential of the HL-LHC to either verify or exclude 
the anomalous $Zb\bar{b}$ couplings observed at LEP through measuring the $Zh$ production rate 
at the HL-LHC, and this conclusion is not sensitive to possible new physics contribution induced by top quark or Higgs boson anomalous couplings in the loop. 
\end{abstract}

\maketitle

\noindent {\bf Introduction:~}
The LEP and SLC experiments have  measured the $Z$ boson couplings and found most of the electroweak data are consistent with the standard model (SM) predictions with a remarkable precision~\cite{ALEPH:2005ab}. However, there are still some experimental results which cannot be explained within the SM framework.  A notorious example is that the bottom quark forward-backward asymmetry ($A_{FB}^b$) measured at the LEP presents a 2.5$\sigma$ deviation with respect to the SM prediction~\cite{ALEPH:2005ab}. As a result, it requires some degree of tuning of the left and right-handed $Zb\bar{b}$ couplings. One class of intriguing models proposed in the literature to explain the puzzling $A_{FB}^b$ data is to allow a sizable right-handed $Zb\bar{b}$ coupling, while keeping the left-handed $Zb\bar{b}$ coupling about the same as the SM value~\cite{Choudhury:2001hs,Agashe:2006at,Liu:2017xmc,Crivellin:2020oup}.
Although such a large discrepancy in $A_{FB}^b$ could be an evidence of new physics (NP) beyond the SM, it is also important to exclude the possibility that it was caused by statistical fluctuation or some subtle systematic errors in experiments.
Resolving this puzzle has became one of the core tasks of the next generation lepton colliders, e.g. CEPC, ILC, CLIC and FCC-ee, which has received much attention by the high energy physics community~\cite{Gomez-Ceballos:2013zzn,Baer:2013cma,Gori:2015nqa,CEPCStudyGroup:2018ghi}.
It has been shown that the $Zb\bar{b}$ anomalous couplings could be well constrained at the future lepton colliders~\cite{Gori:2015nqa}. 
However, a direct measurement of the $Zb\bar{b}$ couplings at the Large Hadron Collider  (LHC) is often ignored in the literature due to the huge backgrounds for detecting the $Z$ boson decaying into a bottom quark and antiquark pair, {\it i.e.,} $Z\to b\bar{b}$ or $Zb$ associated production~\cite{Beccaria:2012xw,Beccaria:2013yya}.

\begin{figure}
	\includegraphics[scale=0.25]{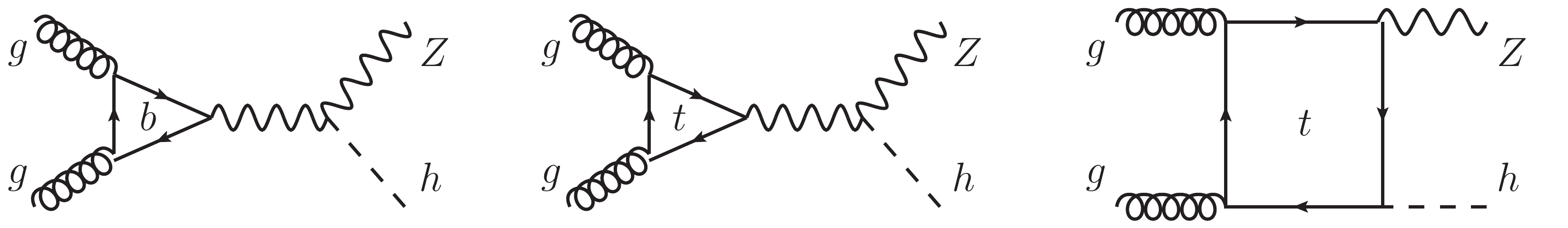}
	\caption{Illustrative Feynman diagrams of $gg\to Zh$ production at the LHC.}
	\label{Fig:fey}
\end{figure}

In this Letter, we propose a novel method to probe the $Zb\bar{b}$ couplings through the associated production of $Z$ and Higgs boson ($h$) via $gg\to Zh$ at the LHC. The $Zb\bar{b}$ couplings contribute to the $Zh$ associated production through bottom quark loop effects in gluon fusion channel, cf.  Fig.~\ref{Fig:fey}. This process has been widely used to 
constrain the top quark anomalous couplings, e.g. $Zt\bar{t}$, $ht\bar{t}$, and it has been shown to be sensitive to many NP effects~\cite{Harlander:2013mla,Hespel:2015zea,Englert:2013vua,Carpenter:2016mwd,Englert:2016hvy,Azatov:2016xik,Goncalves:2018fvn,Harlander:2018yns,Vryonidou:2018eyv,Degrande:2018fog,Xie:2021xtl}. 
For the first time, we demonstrate that this process can also be used to constrain the bottom quark anomalous couplings and to resolve the $A_{FB}^b$ puzzle. 

Owing to charge conjugation invariance, the $Z$-boson couples only axially to the internal quarks in the loop of diagrams shown in Fig.~\ref{Fig:fey}, so that the contribution from a mass-degenerate weak doublet of quarks vanishes. 
It is worthwhile noting that this conclusion will not be influenced by higher order QCD corrections, because QCD theory preserves vector current conservation due to the symmetry of parity~\cite{Hasselhuhn:2016rqt,Davies:2020drs,Chen:2020gae,Alasfar:2021ppe}.
Such property leads to the conclusion that the $gg\to Zh$ production in the SM would only be sensitive to physics of the third generation quarks, i.e, the bottom and top quarks.  Furthermore, as to be shown below, the contribution from the bottom quark is comparable to the top quark in $gg\to Zh$ production. Therefore, such process could be used to probe the axial-vector component of the $Zb\bar{b}$ interaction at hadron colliders. 

The main difficulty of measuring the $Zb\bar{b}$ couplings at the LHC via  $gg\to Zh$ process comes from the contamination of the  top quark contribution in the loop. One can combine the other measurements at the LHC to constrain the top quark anomalous couplings, e.g. $Zt\bar{t}$ and $ht\bar{t}$ couplings~\cite{Cao:2015qta,Cao:2016wib,Bylund:2016phk,Cao:2019ygh,Li:2019uyy,Cao:2020npb}. In this letter, we demonstrate that one could determine the $Zb\bar{b}$ couplings through detecting $Zh$ associated production at the LHC, and the results are not sensitive to  the top quark nor the Higgs boson anomalous couplings. Furthermore, we show that the implication of the $A_{FB}^b$ data at LEP can either be verified or excluded if the central value of the signal strength is found to be less than what SM predicts at the High luminosity LHC (HL-LHC), a proton-proton collider to operate at a center-of-mass energy of $14~{\rm TeV}$ with an integrated luminosity of $3~{\rm ab}^{-1}$. In that case, our method can also resolve the degeneracy of the $Zb\bar{b}$ couplings, presently allowed by the precision electroweak data at LEP and SLC. 

\vspace{3mm}
\noindent {\bf $Zh$ production via gluon fusion:~}%
We consider the following effective Lagrangian related to $Zh$ associated production,
\begin{align}
\mathcal{L}&=\frac{g_W}{2c_W}\bar{b}\gamma_\mu(\kappa_v^b v_b^{\rm SM}-\kappa_a^ba_b^{\rm SM}\gamma_5) bZ_\mu+\frac{m_Z^2}{v}\kappa_ZhZ_{\mu}Z^{\mu}\nn\\
&
+\frac{g_W}{2c_W}\bar{t}\gamma_\mu(\kappa_v^t v_t^{\rm SM}-\kappa_a^ta_t^{\rm SM}\gamma_5) tZ_\mu
-\frac{m_t}{v}\kappa_t\bar{t}th,
\label{eq:one}
\end{align}
where $g_W$ is the weak gauge coupling, $c_W$ is the cosine of the weak mixing angle $\theta_W$ and $v=246~{\rm GeV}$ is the Higgs vacuum expectation value. The gauge coupling strength modifiers $\kappa_{v,a}^{b,t}$ and $\kappa_{t,Z}$ are introduced to include possible NP effects. The vector and axial-vector couplings of $Z$ boson to bottom ($b$) and top ($t$) quarks in the SM are $v_q^{\rm SM}=T_3-2Qs_W^2$ and $a_q^{\rm SM}=T_3$, where $(T_3,Q)=(1/2,2/3)$ and $(-1/2,-1/3)$ for $t$ and $b$, respectively, with $s_W\equiv\sin\theta_W$. We calculate the helicity amplitudes 
$M_{\lambda_1,\lambda_2,\lambda_3}$ of the channel $g(\lambda_1)g(\lambda_2)\to Z(\lambda_3)h$ using FeynArts~\cite{Hahn:2000kx} and FeynCalc~\cite{Shtabovenko:2016sxi}, where $\lambda_i=+,-,0$ labels the helicity of particle $i$. 
Below, we show the explicit expression of the dominant helicity amplitudes, which can be written, separately for triangle ($\triangle$) and box ($\square$) diagrams, as
\begin{align}
M_{++0}^{\triangle}&=2\frac{\sqrt{\lambda}}{m_Z}\sum_{t,b}\left[\kappa_a^q\kappa_Z\frac{a_{q}^{\rm SM}g_{hZZ}}{m_Z^2}\left(F_\triangle(s,m_q^2)+2\right)\right]N,\nn\\
M_{++0}^{\square} &=-\frac{4}{m_Z\sqrt{\lambda}}\kappa_a^t\kappa_tg_{htt}a_t^{\rm SM}m_t\left[F_{++}^0+(t\leftrightarrow u)\right]N,
\label{eq:hel}
\end{align}
where
\beq
\lambda=s^2+m_Z^4+m_h^4-2(sm_Z^2+m_Z^2m_h^2+m_h^2s),\quad N=\frac{\alpha_s g_W}{32\pi c_W},
\eeq 
and $s, t, u$ are the usual Mandelstam variables for describing the scattering of $gg \to Zh$. The other couplings are defined as $g_{hZZ}=2m_Z^2/v$ and $g_{htt}=-m_t/v$. 
The helicity amplitudes $M_{--0}^{\triangle,\square}=-M_{++0}^{\triangle,\square}$, while 
the other helicity configurations could be ignored due to their small numerical contributions, about 0.1\% at the 14 TeV LHC, to the inclusive production cross section.
The definition of the scalar functions $F_\triangle$ and $F_{++}^0$ in Eq.~(\ref{eq:hel}) could be found in Ref.~\cite{Kniehl:2011aa}. We have compared our analytical results to  MadGraph5~\cite{Alwall:2014hca} and found perfect agreement.

\begin{figure}
	\includegraphics[scale=0.24]{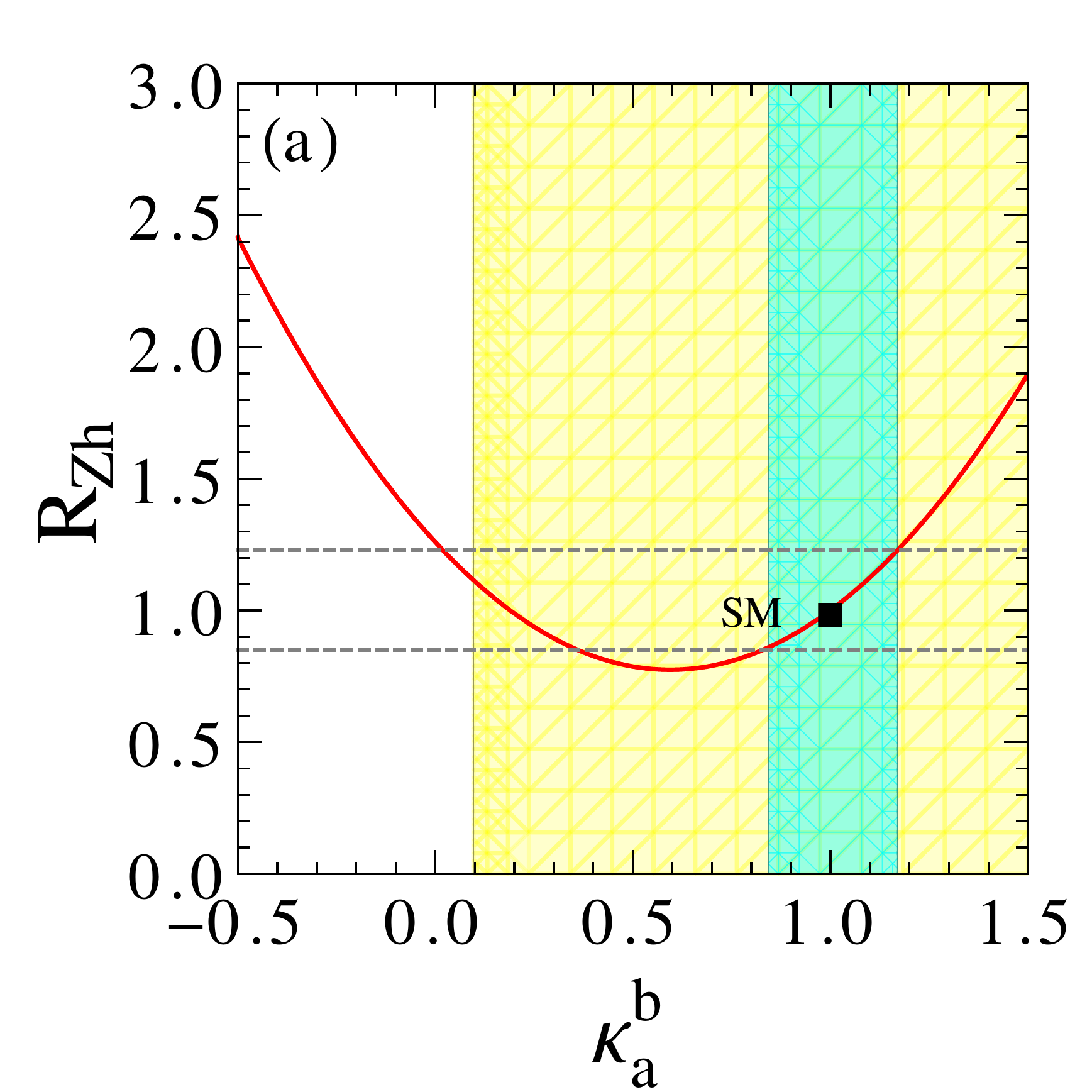}
	\includegraphics[scale=0.24]{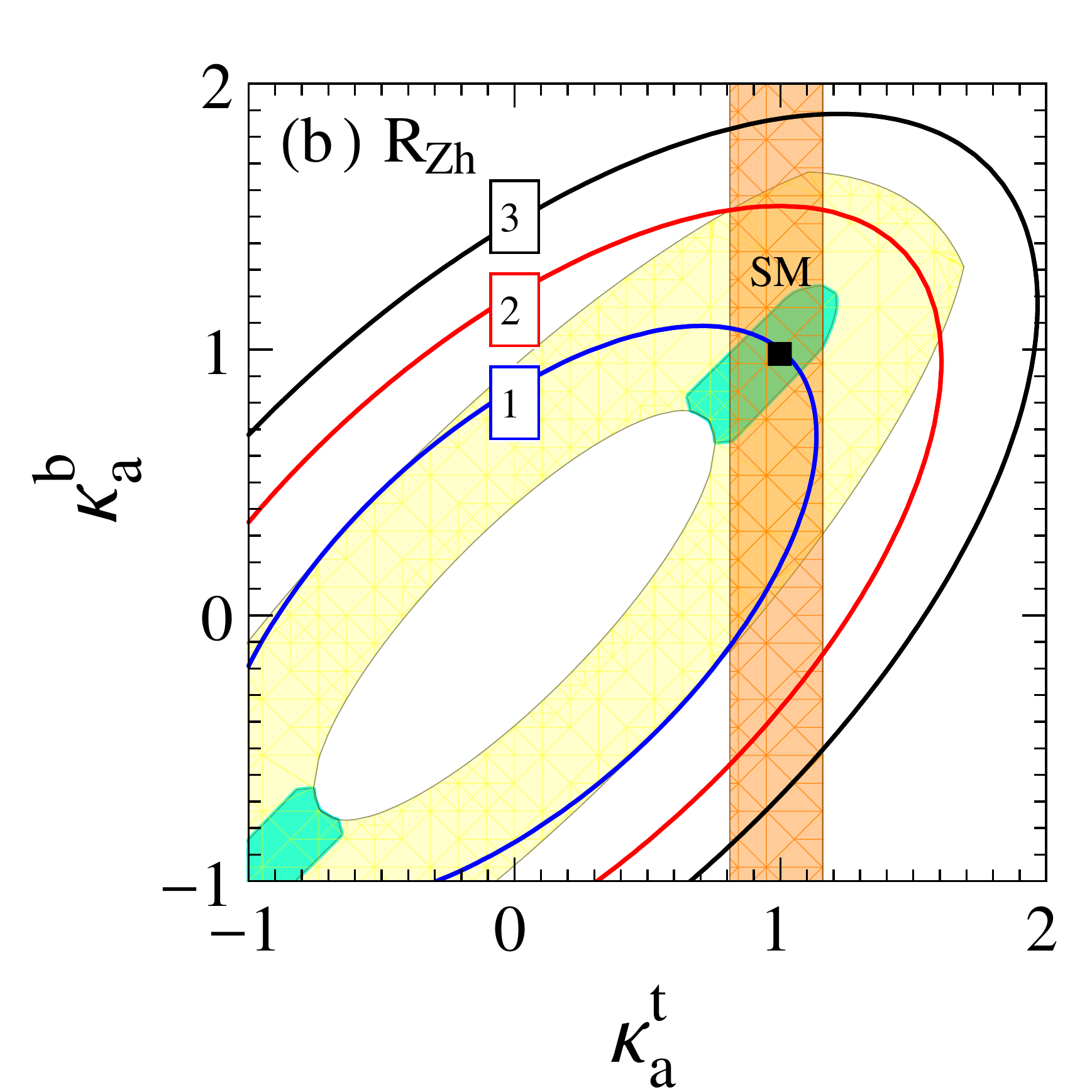}
	\includegraphics[scale=0.24]{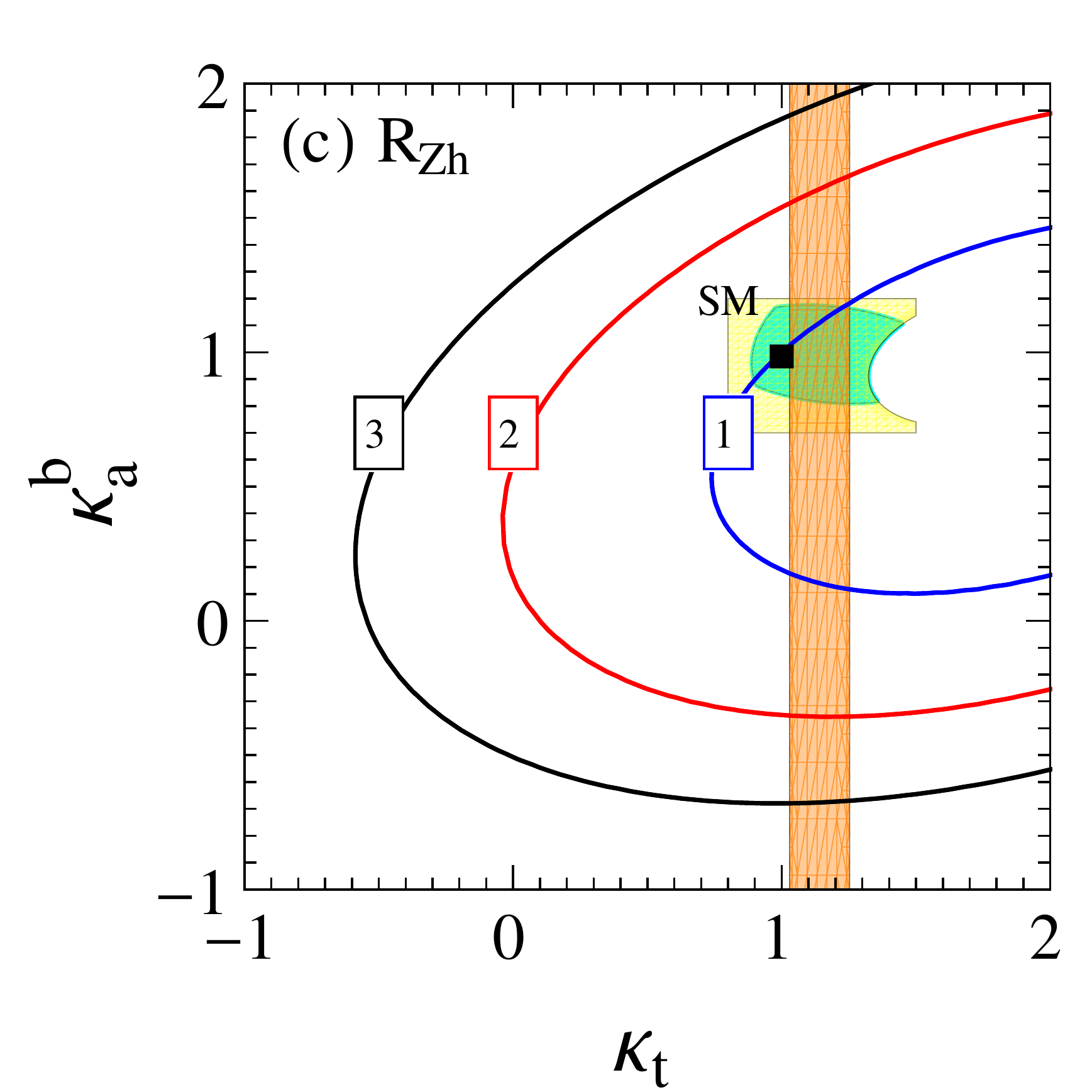}
	\includegraphics[scale=0.24]{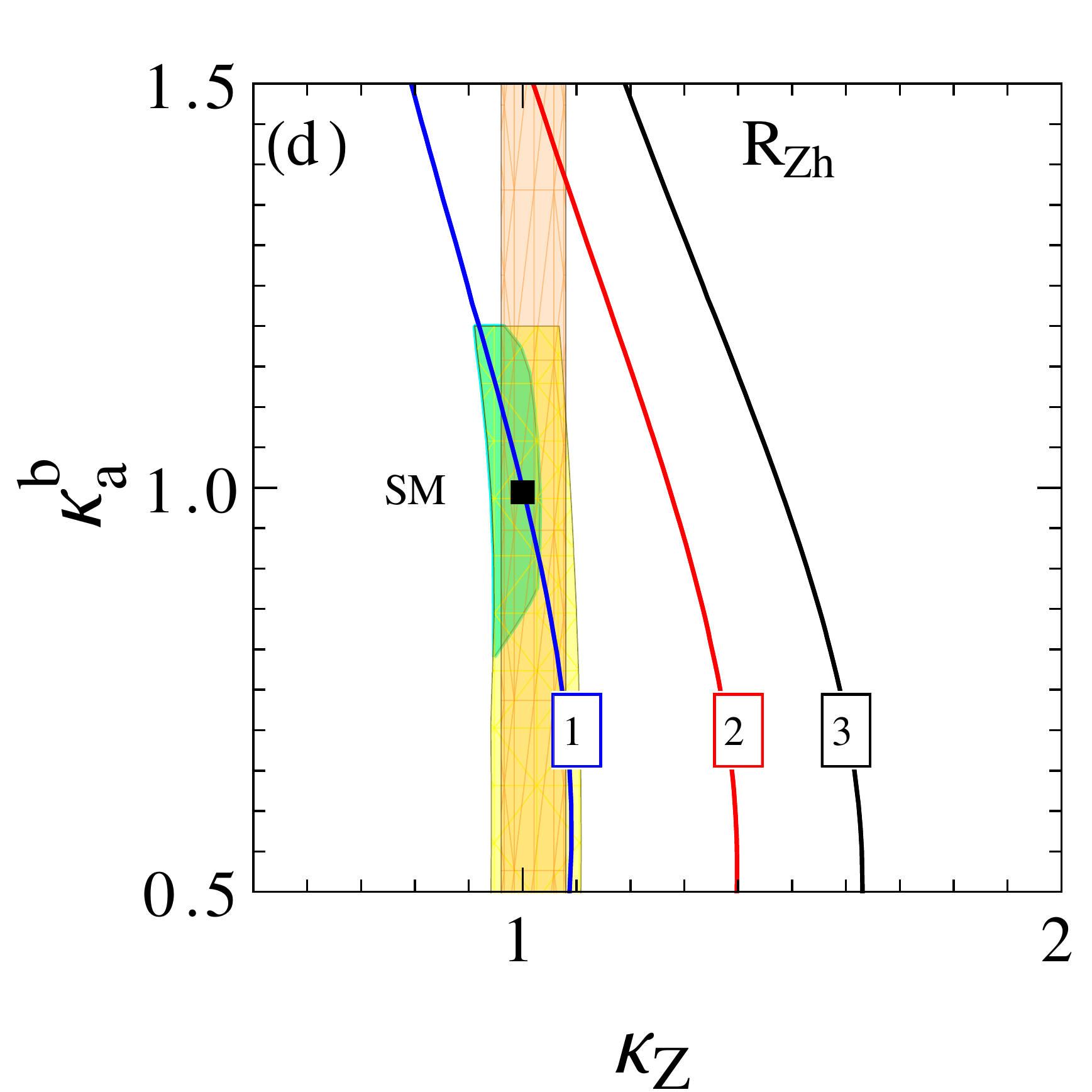}
	\caption{The contours of $R_{Zh}$ in the plane of anomalous couplings at the 14 TeV LHC. The cyan region denotes the constraints, at $1\sigma$ level, provided by the measurements of $pp\to Zh$ at the 13 TeV LHC, while the yellow shaded region denotes the impact after removing the two high $p_T^Z$ data, depicted in Fig. 4 of Ref.~\cite{Aad:2020eiv}.  The orange bands in (b), (c) and (d) come from the constraints of $Zt\bar{t}$, $ht\bar{t}$ and $hZZ$ coupling measurements at the 13 TeV LHC, respectively. }
	\label{Fig:Ratio}
\end{figure}

\begin{figure}
	\includegraphics[scale=0.24]{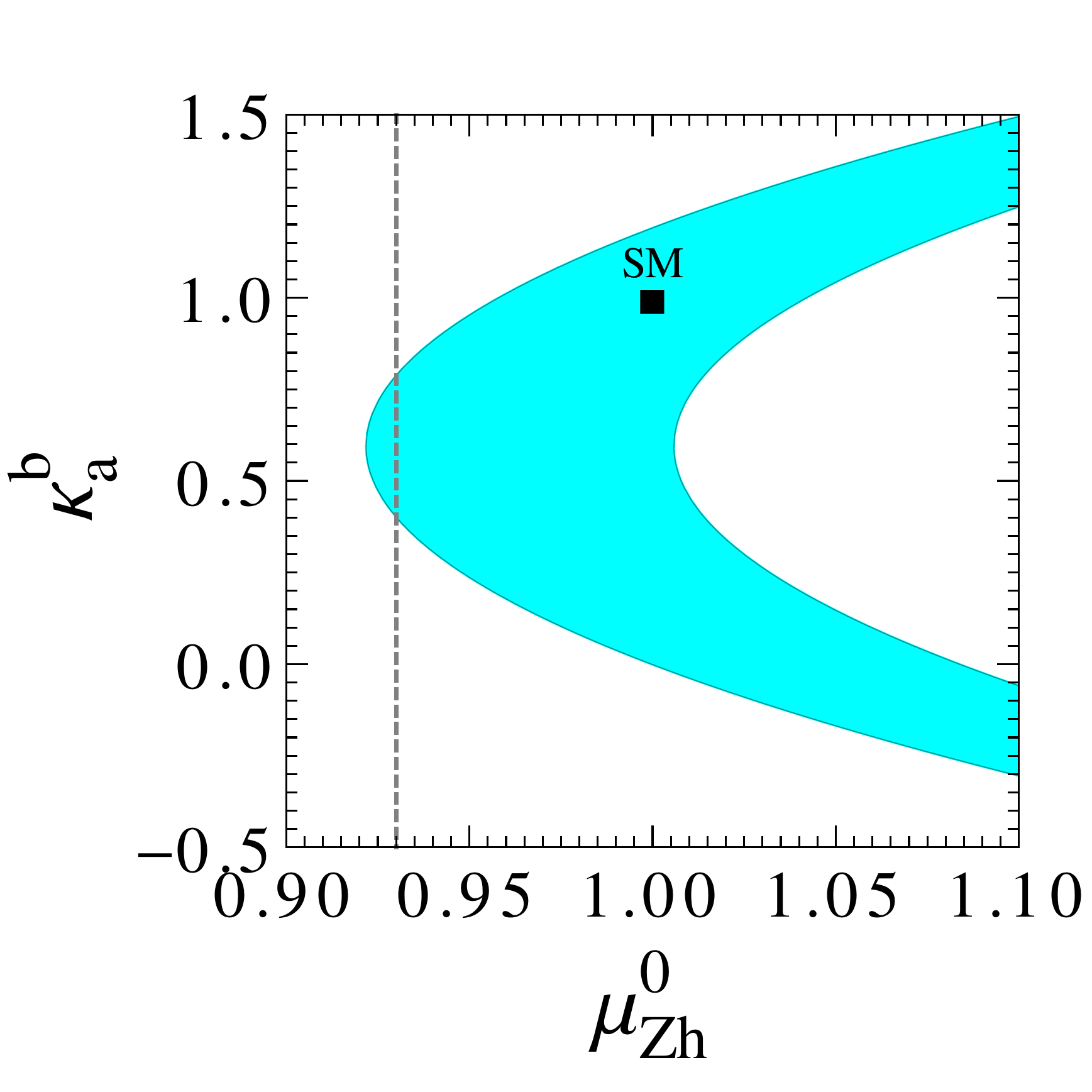}
	\caption{Expected uncertainty for measuring $\kappa_a^b$ via $Zh$ associated production at the HL-LHC, as a function of the signal strength $\mu_{Zh}^0$.
		The vertical dashed line indicates $\mu_{Zh}^0=0.93$.
}
	\label{Fig:HL1}
\end{figure}

\begin{figure}
	\includegraphics[scale=0.24]{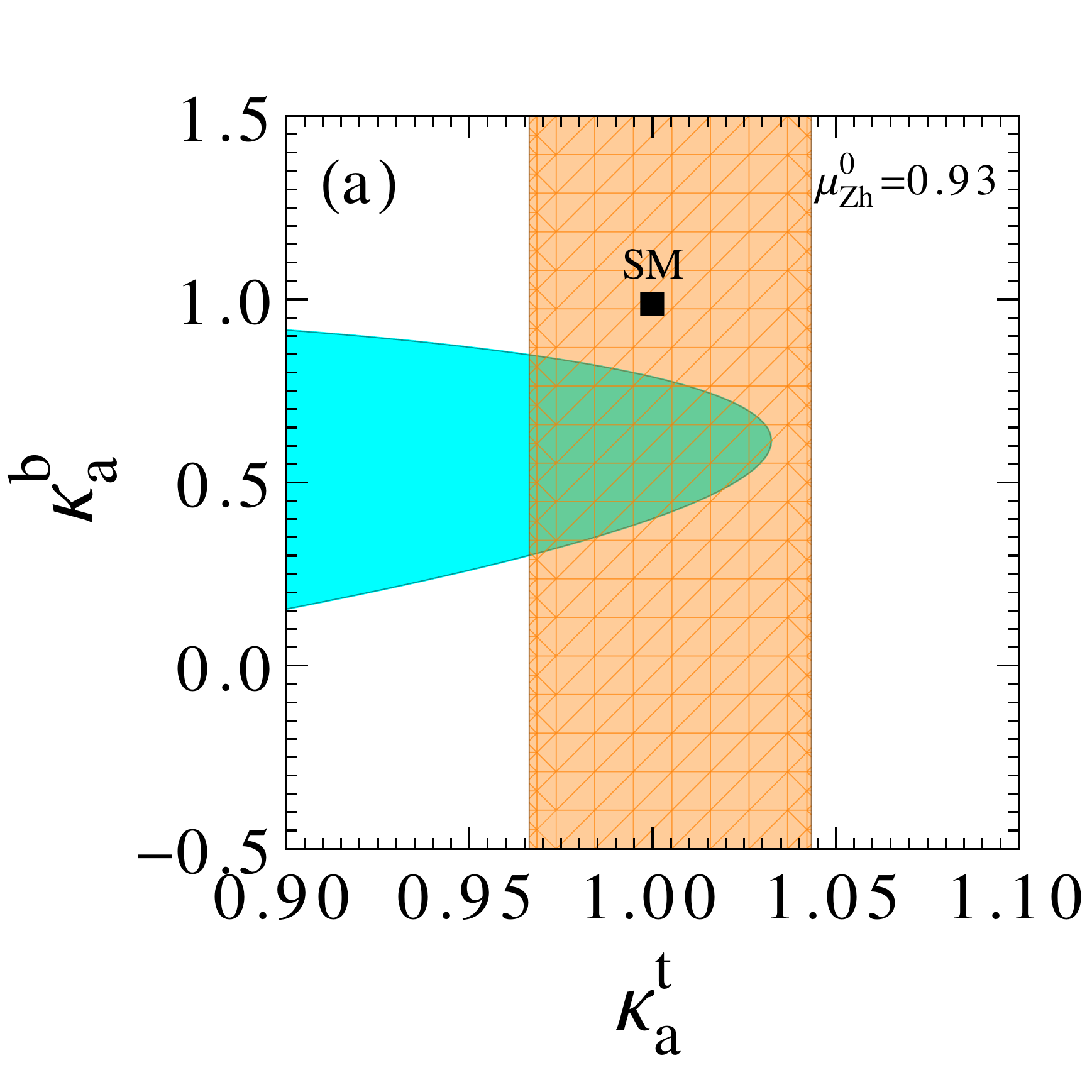}
	\includegraphics[scale=0.24]{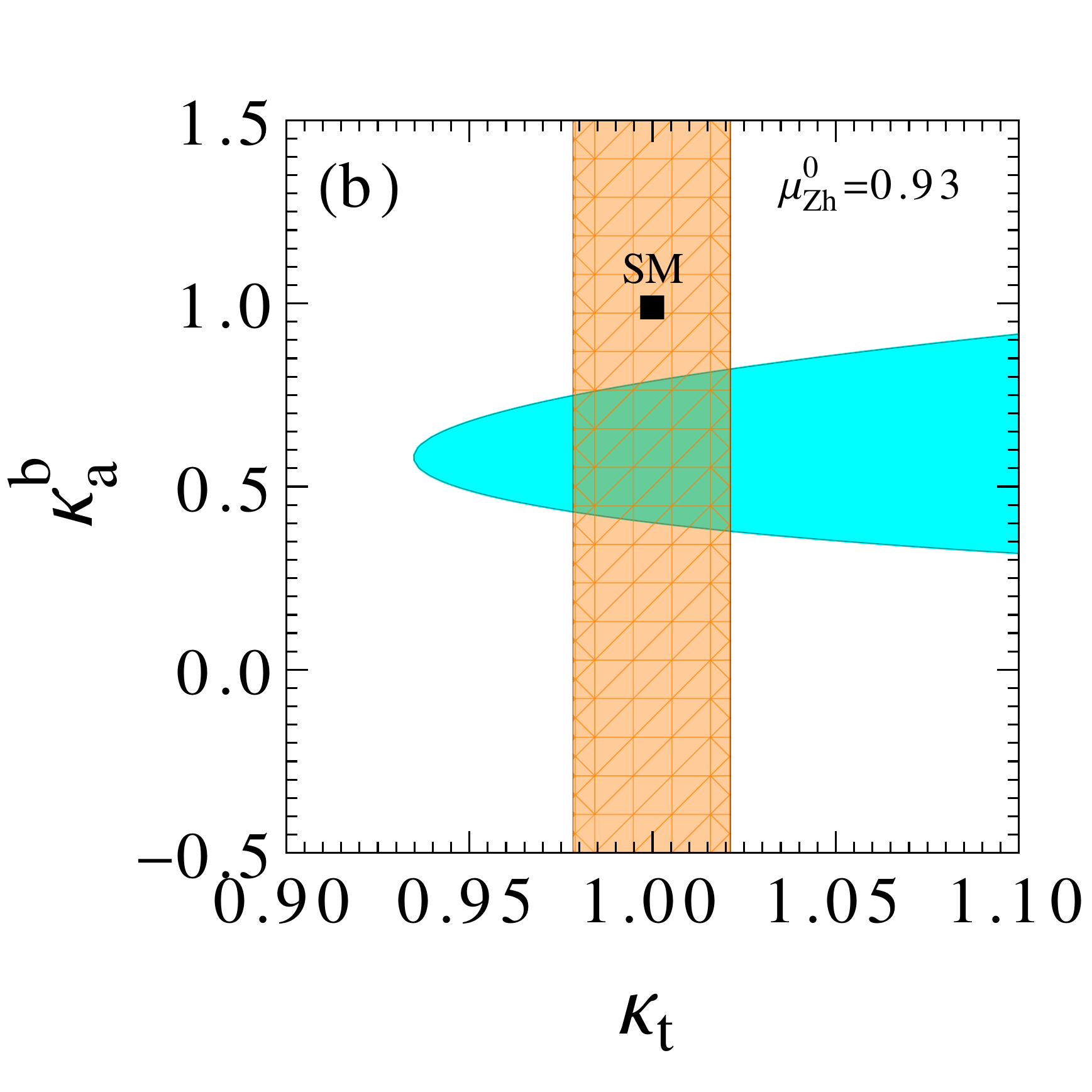}
	\includegraphics[scale=0.24]{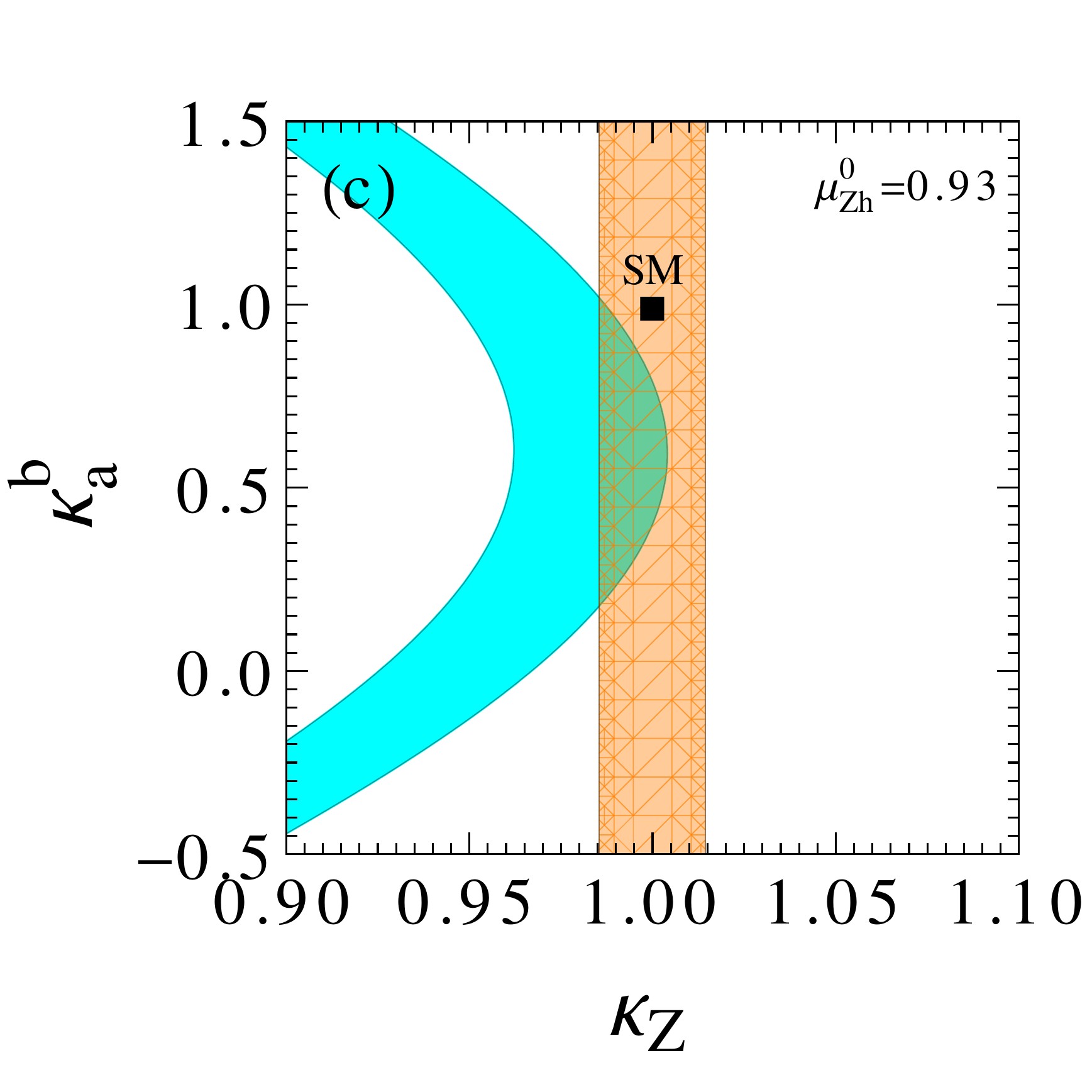}
	\caption{The blue bands represent the expected uncertainty for measuring $\kappa_a^b$ via $Zh$ associated production at the HL-LHC. The central value of signal strength is assumed to be $\mu_{Zh}^0=0.93$.  The orange bands in (a), (b) and (c) represent the expected $1\sigma$ constraints provided by the $Zt\bar{t}$, $ht\bar{t}$ and $hZZ$ coupling measurements at the HL-LHC, respectively.
 }
	\label{Fig:HL1b}
\end{figure}

\vspace{3mm}
\noindent {\bf Sensitivity at the LHC:~}%
Below, we consider the impact of the non-standard $Zb\bar{b}$, $Zt\bar{t}$, $ht\bar{t}$ and $hZZ$ couplings to the inclusive cross section of $gg\to Zh$ at the 14 TeV LHC. In order to compare $\sigma(gg\to Zh)$ with the SM prediction, we define a ratio $R_{Zh}\equiv\sigma(gg\to Zh)/\sigma(gg\to Zh)^{\rm SM}$.
Figure~\ref{Fig:Ratio} displays the contours of $R_{Zh}$ in the plane of anomalous couplings. The cyan region denotes the constraints from 
the measurements of inclusive cross section and transverse momentum distribution of $Z$ boson ($p_T^Z$) in the $pp \to Zh$ production at the 13 TeV LHC~\cite{Aaboud:2018zhk,Aaboud:2019nan,Aad:2020jym,Aad:2020eiv,CMS:2020gsy}. 
The dominant constraint on the parameter space comes from the ATLAS $Zh$ data with high $p_T^Z$~\cite{Aad:2020eiv}, which shows a large deviation from the central value of the signal strength ($\mu_{Zh}\equiv\sigma(pp\to Zh)/\sigma(pp\to Zh)^{\rm SM}$) predicted by the SM, i.e. $\mu_{Zh}=0.34^{+0.75}_{-0.7}$ and $\mu_{Zh}=0.28^{+0.97}_{-0.83}$ for $p_T^Z\in [250,400]~{\rm GeV}$ and $p_T^Z>400~{\rm GeV}$, respectively, cf. Fig.4 of Ref.~\cite{Aad:2020eiv}.
If we exclude those two high $p_T^Z$ data points, whose central values happen to be much smaller than the SM predictions, 
the allowed parameter space is depicted as the yellow band in Fig.~\ref{Fig:Ratio}(a), which shows a much wider uncertainty in $\kappa_a^b$ as compared to the cyan band (with $0.84<\kappa_a^b<1.17$ at $1\sigma$ level). 
Hence, it is important to improve the measurement of the $p_T^Z$ distribution of $Zh$ production at the HL-LHC. 
The higher order QCD correction effects are taken into account by introducing a constant $k$-factor for both $q\bar{q}\to Zh$ and $gg\to Zh$ production processes in the analysis, with  $k_{qq}=1.3$ and $k_{gg}=2.7$~\cite{deFlorian:2016spz,Hasselhuhn:2016rqt}.
The orange bands in Fig.~\ref{Fig:Ratio}(b)-(d) show the constraints imposed by the  measurements of $Zt\bar{t}$, $ht\bar{t}$ and $hZZ$ couplings at the 13 TeV LHC~\cite{Sirunyan:2018zgs,Aaboud:2018urx,Sirunyan:2018hoz,CMS:2019too,Aaboud:2019njj,Aad:2020wog,Aad:2020mkp}, respectively. 
They were obtained by analyzing the production of $Zt\bar{t}$, $Ztj$, $ht\bar{t}$ and $gg \to h \to ZZ^*$, etc. 
The cyran region shown in Fig.~\ref{Fig:Ratio}(b) is constrained by the 13 TeV $Zh$ data while letting both $\kappa_a^t$ and $\kappa_v^t$ freely vary within the allowed range imposed by the measurement of $Zt\bar{t}$ couplings. This yields only a slightly larger uncertainty band as  $0.66<\kappa_a^b<1.23$. 
Similar analyses, but separately for the anomalous couplings $\kappa_t$ and $\kappa_Z$, are shown in Fig.~\ref{Fig:Ratio}(c) and (d).
Within the constraints from the current data, the cross section $\sigma(gg\to Zh)$ could 
differ from the SM prediction by about 
$20\%\sim 30\%$ (or 100\%, if without the inclusion of two above-mentioned high $p_T^Z$ data points) which is a large enough deviation that can be detected at the HL-LHC~\cite{Cepeda:2019klc}.

At the HL-LHC, many experimental measurements can be further improved as compared to the present data. The error of the signal strengths of $Zh$ and $ht\bar{t}$ productions can be reduced to about 4.2\% and 4.3\%, respectively, while the uncertainty of the branching ratio of $h\to ZZ^*$ would be at 2.9\%~\cite{Cepeda:2019klc}. The limit on $\kappa_a^t$ will also be highly improved through measuring the $t\bar{t}h$, $thj$ and $gg\to ZZ$ productions~\cite{Azatov:2016xik,Cao:2020npb}. We summarize the expected constraints from the above production processes at the HL-LHC in Figs.~\ref{Fig:HL1} and \ref{Fig:HL1b}.
Fig.~\ref{Fig:HL1} shows that the expected precision for measuring $\kappa_a^b$ at the HL-LHC is sensitive to the central value of the signal strength $\mu_{Zh}^0$.  Taking  $\mu_{Zh}^0=0.93$ as an illustration, 
$\kappa_a^b$ will be constrained to be [0.40,0.78] at the $1\sigma$ level  when we consider $\kappa_a^b$ alone.
As shown in Fig.~\ref{Fig:fey}, 
the $Zh$ associated production cross section also depends on other parameters in Eq.~(\ref{eq:one}).
Fig.~\ref{Fig:HL1b} shows the results when we consider only two parameters at a time~\cite{Cepeda:2019klc}. 
The blue band represents the $1\sigma$ uncertainty of the 
$Zh$ associated production cross section measurement with 
$\mu_{Zh}^0=0.93$. 
In the same figure, 
the orange shaded regions show the expected $1\sigma$ constraints provided by the $Zt\bar{t}$, $ht\bar{t}$ and $hZZ$ coupling measurements at the HL-LHC, respectively.   
It shows that the HL-LHC measurements of the other processes, such as 
$Zt\bar{t}$, $ht\bar{t}$ and $hZZ$, will mainly constrain all the anomalous couplings
in Eq.~(\ref{eq:one}) except for  $\kappa_a^b$. Hence, any substantial deviation
observed in $\mu_{Zh}$ would be ascribed to $\kappa_a^b$,  whose allowed range will not be noticeably modified by a combined fit with the inclusion of the above-mentioned  processes.

\begin{figure}
	\includegraphics[scale=0.24]{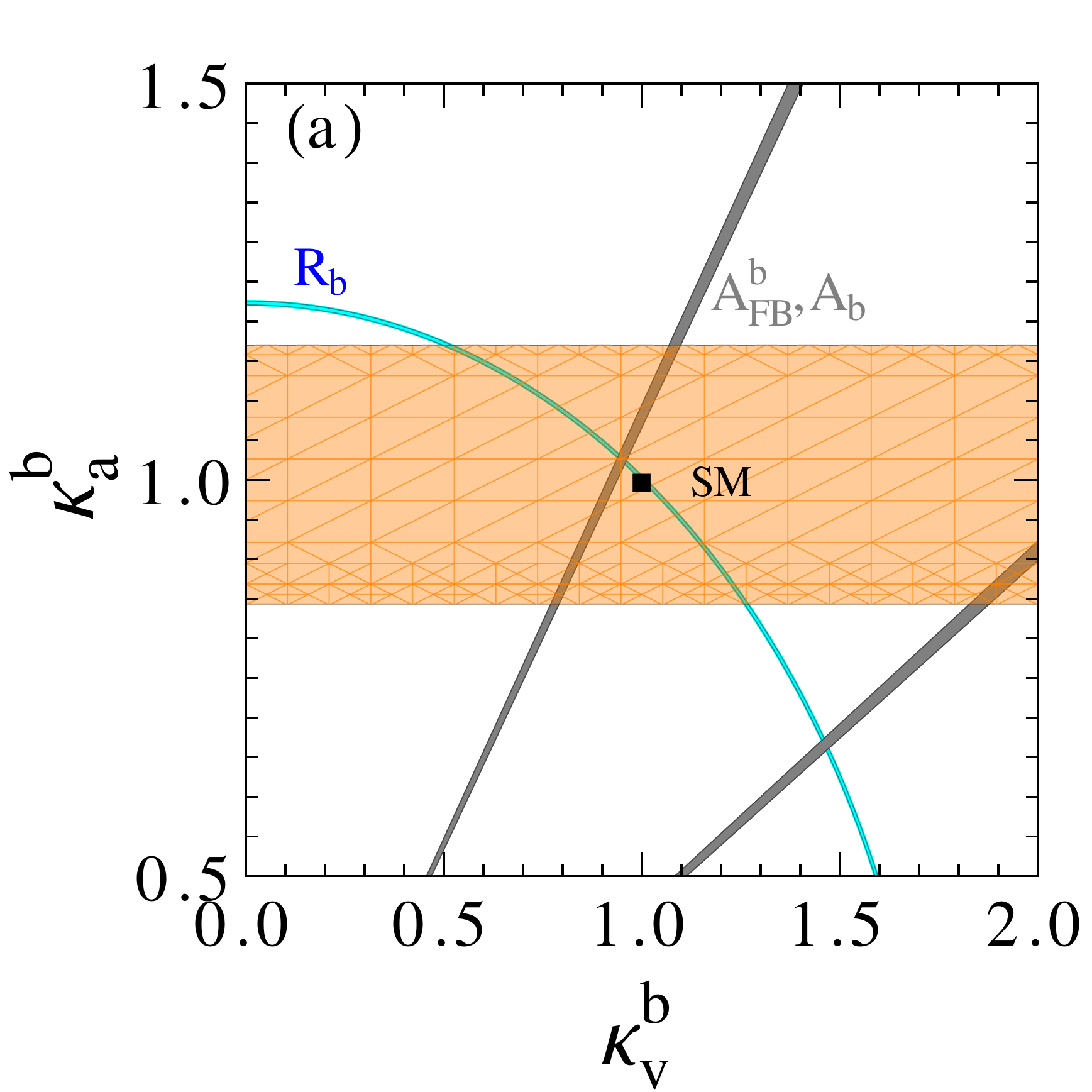}
	\includegraphics[scale=0.24]{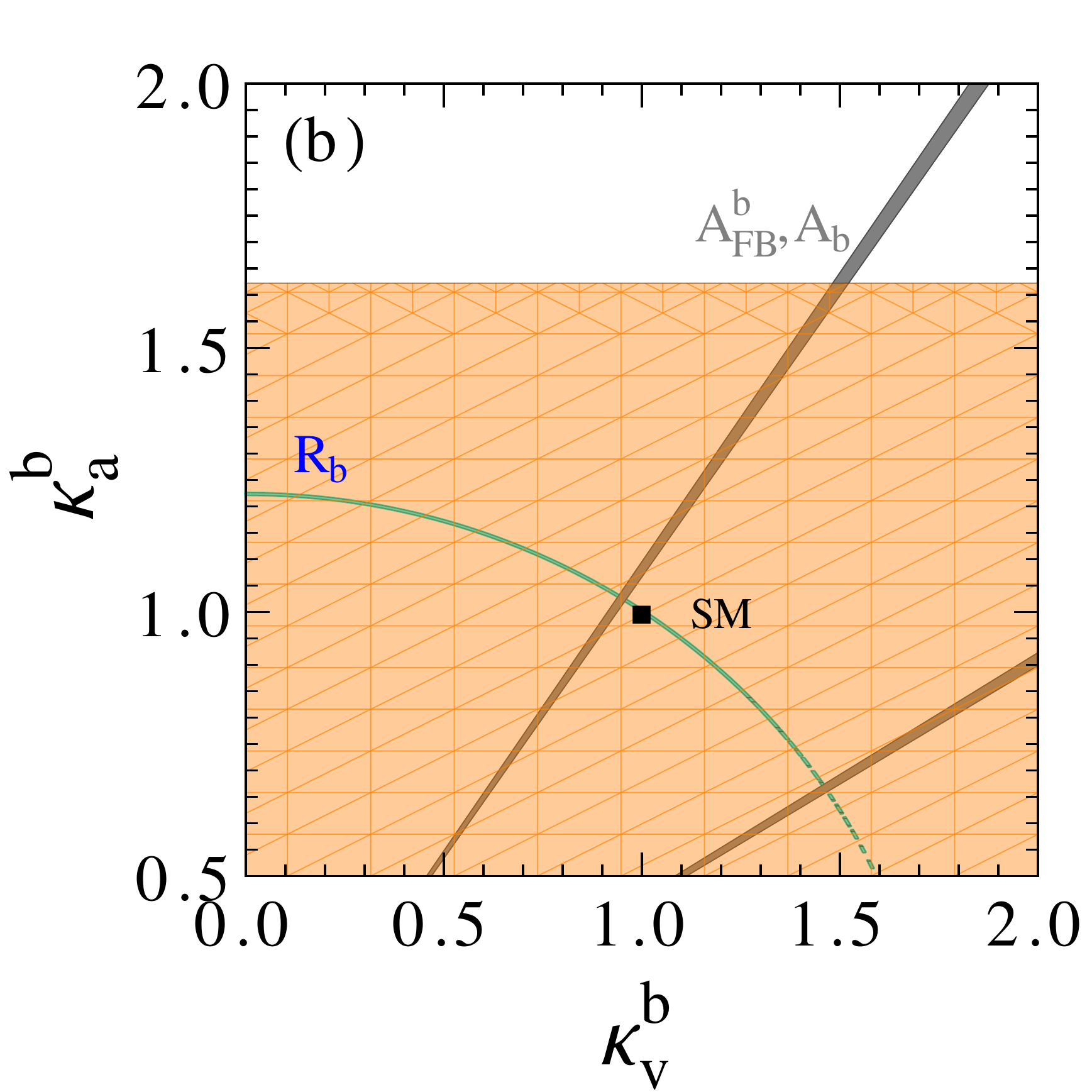}
	\caption{Present constraints on the axial-vector $Zb\bar{b}$ coupling $\kappa_a^b$. The cyan and gray regions come from the $R_b$, $A_{FB}^b$, and $A_b$ measurements at LEP and SLC, respectively. The orange band in (a) comes from the measurements of inclusive cross section and $p_T^Z$ distribution of $Zh$ associated production at the 13 TeV LHC, while (b) is the result after removing the two high $p_T^Z$ data in Fig. 4 of Ref.~\cite{Aad:2020eiv}.  }
	\label{Fig:zbb}
\end{figure}

\vspace{3mm}
\noindent {\bf Break the $Z\bar{b} b$ coupling degeneracy:~}%
The $Zb\bar{b}$ couplings are also well constrained by the LEP and SLC electroweak data and mainly determined by the $Z$-pole measurements at the LEP: $R_b$ and $A_{FB}^b$.  The $R_b$ is defined by the ratio, $R_b=\Gamma(Z\to b\bar{b})/\sum_q\Gamma(Z\to q \bar{q})$,
where the sum in the denominator includes all quarks except the top quark. The $R_b$ measurement agrees with the SM prediction very well, while the $A_{\rm FB}^b$ at the $Z$-pole exhibits  a large deviation from the SM prediction with a significance around $2.5\sigma$~\cite{ALEPH:2005ab,Baak:2014ora}. The SLC with polarized beam can directly probe the bottom quark asymmetry $A_b$ which was found to be consistent with the SM within $1\sigma$~\cite{ALEPH:2005ab}.  As pointed out in Ref.~\cite{Choudhury:2001hs}, the $Z$-pole data alone can not fully determine the $Zb\bar{b}$ interactions due to the sign ambiguities of the couplings. With the help of $A_{FB}^b$ measurements conducted in the off-shell $Z$ boson region, the part of parameter space with $\kappa_{a,v}^b<0$ has been excluded. However, it remains to be difficult to resolve the apparent  degeneracy in the parameter space with  $\kappa_{a,v}^b>0$ due to the limited statistics for data away from the $Z$-pole.
As shown in Fig.~\ref{Fig:zbb}, both sets of $\kappa_a^b$ and $\kappa_v^b$ values, with $(\kappa_a^b,\kappa_v^b)=(1.03, 0.95)$ or  $(0.67,1.46)$, are consistent with the precision electroweak data at LEP and SLC.

\begin{figure}
	\includegraphics[scale=0.24]{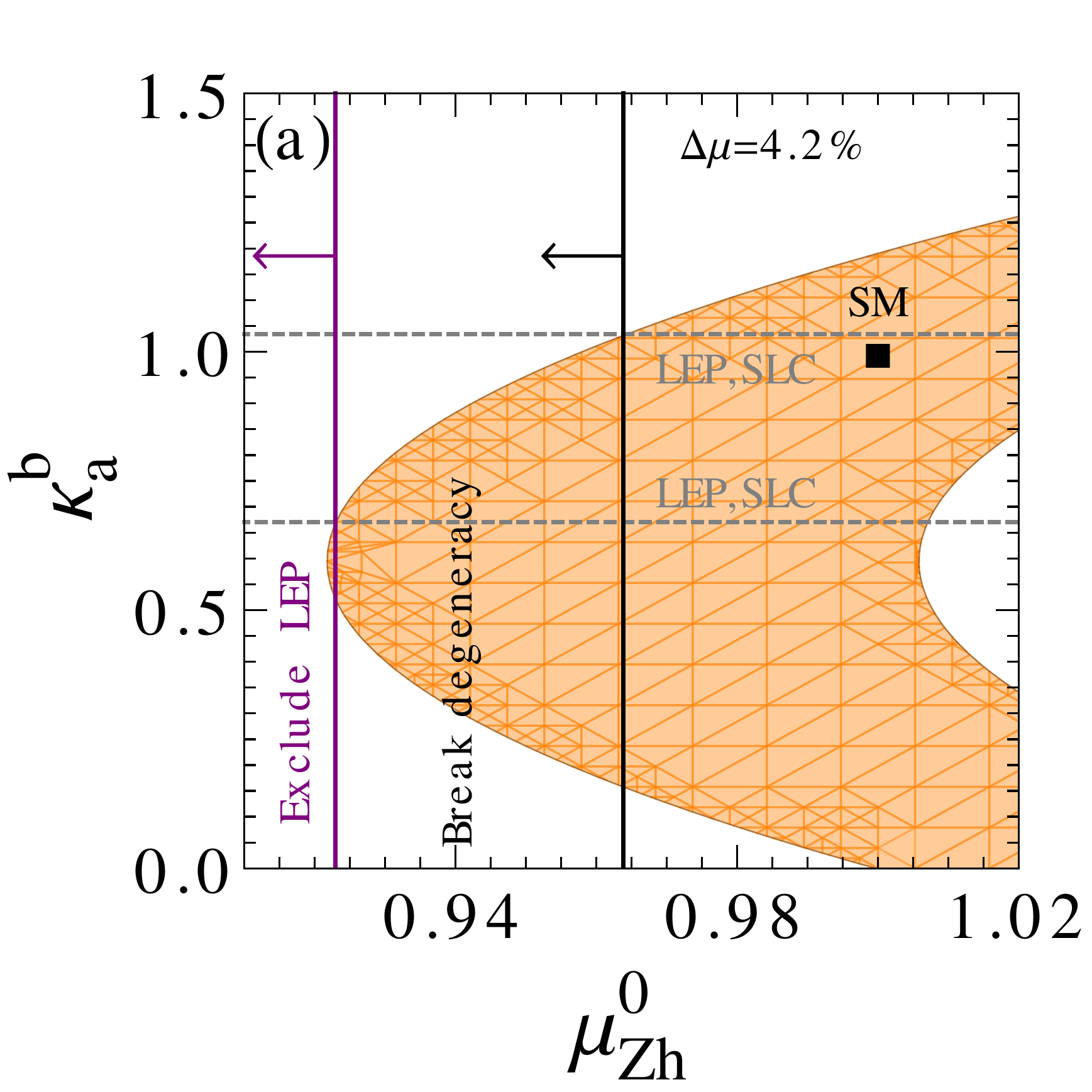}
	\includegraphics[scale=0.24]{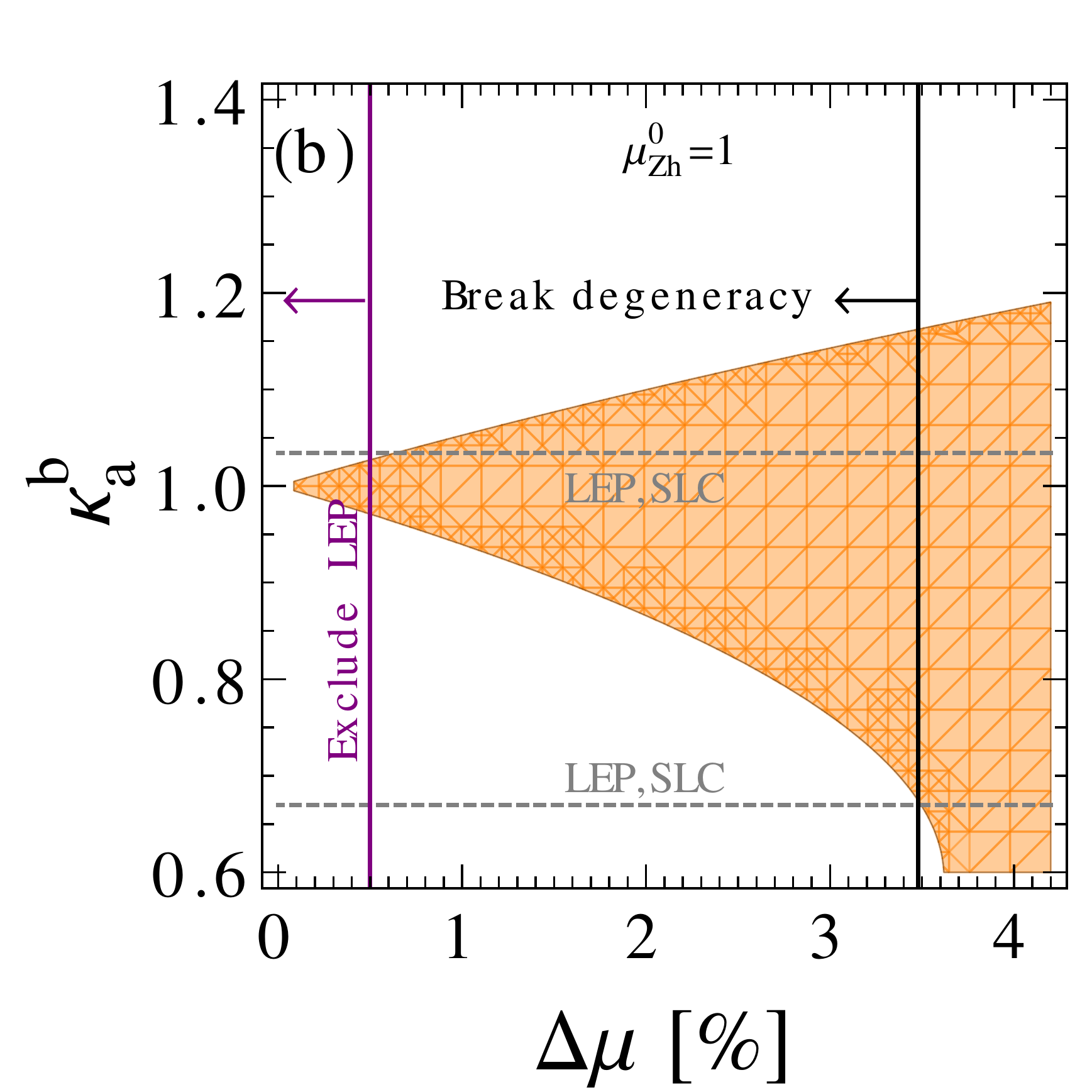}
	\caption{Expected signal strength $\mu_{Zh}^0$ (a) and precision $\Delta\mu$ (b) for breaking the $Zb\bar{b}$ coupling degeneracy  (black perpendicular line) and excluding LEP $A_{FB}^b$ measurement (purple perpendicular line) at the 14 TeV LHC. 
	The two gray (horizontal) dashed lines indicate the degeneracy (at $\kappa_a^b$= 1.03 and 0.67) found in the analysis of LEP and SLC data.
	The orange band in (a) shows the expected uncertainty on the determination of $\kappa_a^b$
	from measuring the inclusive cross section of the $Zh$ production at the HL-LHC with a fixed $\Delta\mu=4.2\%$, while (b) is the similar result but with a fixed $\mu_{Zh}^0=1$.   
}
	\label{Fig:zbb_HL}
\end{figure}

To break this degeneracy, one needs to analyze the $Zh$ data collected at LHC. 
In Fig.~\ref{Fig:zbb}, we compare 
the precision on the determination of the axial-vector component of the $Zb\bar{b}$ anomalous coupling via the measurements of inclusive cross section and transverse momentum distribution of $Z$ boson in the $Zh$ production at the 13 TeV LHC to that implied by the precision electroweak data at LEP and SLC. Here, we focus on the parameter space with $\kappa_{a,v}^b>0$.
The cyan and gray shaded regions denote the constraint from $R_b$ and ($A_{FB}^b$, $A_b$) measurements with $1\sigma$ accuracy, respectively. The orange region in Fig.~\ref{Fig:zbb}(a) is consistent with the current $Zh$ production measurements, while the orange band in Fig.~\ref{Fig:zbb}(b) shows the allowed parameter space region after we remove the above-mentioned two high $p_T^Z$ data points~\cite{Aad:2020eiv}. 
It appears that the current measurement of the $Zh$ inclusive cross section at the LHC has broken the degeneracy in the allowed $\kappa_a^b$ and $\kappa_v^b$ values implied by the precision electroweak data, resulting the preferred values of $\kappa_a^b$ and $\kappa_v^b$ to be close to 1, the SM values. 
However, the degeneracy would remain after removing the two high $p_T^Z$ data points whose central values are in conflict with the SM predictions. 
Hence, it is important to have a more precise  measurement of the $p_T^Z$ distribution in the $Zh$ events at the LHC.

Next, we discuss the potential of the HL-LHC to break the above-mentioned degeneracy and determine the value of $\kappa_a^b$.
The expected constraint on $\kappa_a^b$ derived from the $Zh$ inclusive cross section measurement at the HL-LHC, assuming an projected uncertainty $\Delta\mu=4.2\%$~\cite{Cepeda:2019klc}, is shown as the 
orange band in Fig.~\ref{Fig:zbb_HL}(a), in which the two horizontal lines indicate the two values (1.03 and 0.67) of $\kappa_a^b$ consistent with the precision electroweak data at LEP and SLC.  
The degeneracy of $\kappa_a^b$ (and $\kappa_v^b$) found in interpreting the precision electroweak data at LEP and SLC can be broken by 
the measurement of $Zh$ production cross section at HL-LHC if $\mu_{Zh}^0$ is measured to be less than 0.964 (indicated by the black perpendicular line), which excludes the solution of $\kappa_a^b=1.03$
allowed by the precision electroweak data and implies new physics beyond the SM must exist (for $\mu_{Zh}^0 \neq 1$).
In case that $\mu_{Zh}^0$ is measured to be less than 0.923 (indicated by the purple perpendicular line), the $Zh$ cross section measurement at HL-LHC would exclude the interpretation of the precision electroweak data at LEP and SLC by introducing merely the anomalous $\kappa_a^b$  and $\kappa_v^b$ couplings.
Moreover, in that case, $\kappa_a^b$ can be well determined.

On the other hand, when $\mu_{Zh}^0=1$, it becomes challenging  to test against the $A_{FB}^b$ measurement at LEP by measuring the $Zh$ cross section at HL-LHC, due to the large uncertainty in $\kappa_a^b$, cf. Fig.~\ref{Fig:HL1}. 
To achieve that goal, a much higher precision of the $Zh$ cross section measurement is needed.  Fig.~\ref{Fig:zbb_HL}(b) shows the required precision $\Delta\mu$ to break the apparent degeneracy in the $Z \bar b b$ couplings, as implied by the LEP and SLC electroweak data, is $3.5\%$ (black perpendicular line). 
To exclude
the interpretation of the $A_{FB}^b$ data at LEP by introducing merely the anomalous $\kappa_a^b$  and $\kappa_v^b$ couplings would require $\Delta\mu=0.5\%$ (indicated by the purple perpendicular line).
However, current study on the $Zh$ cross section measurement at HL-LHC ~\cite{Cepeda:2019klc} projects a 
$2.6\%$ statistical error, $1.3\%$ experimental systematic error and $3.1\%$ theoretical uncertainty. Hence, 
$\Delta\mu=3.5\%$ might be hopeful to reach at HL-LHC, with an improvement in the theoretical calculation accuracy. But, it would be very challenging to reach $\Delta\mu=0.5\%$.
Nevertheless, measuring the cross section of $Zh$ production can clearly break the apparent degeneracy of the $Z \bar b b$ couplings, as implied by the precision electroweak data at LEP and SLC.

\vspace{3mm}
\noindent {\bf Conclusions:~}%
In this Letter, we propose a novel signature to probe the anomalous $Zb\bar{b}$ couplings through measuring the cross section of the $Zh$ associated production via gluon-gluon fusion at the LHC. Our method could be used 
to break the apparent degeneracy in the $Z \bar b b$ couplings, as implied by the LEP and SLC precision electroweak data, including the 
long-standing discrepancy of the $A_{FB}^b$ data from LEP. 
We show that the $Zh$ cross section at the one-loop level depends on the axial-vector component of the  $Zb\bar{b}$ coupling.  
The determination of   $\kappa_a^b$ is sensitive to the central value of the signal strength ($\mu_{Zh}^0$) of $Zh$ production,
and is not sensitive to possible new physics contribution induced by top quark or Higgs boson anomalous couplings in the loop.
The HL-LHC measurements of the other processes, such as 
$Zt\bar{t}$, $ht\bar{t}$ and $hZZ$, will mainly constrain all the anomalous couplings
in Eq.~(\ref{eq:one}) except for $\kappa_a^b$. Hence, any substantial deviation
observed in $\mu_{Zh}$ would be ascribed to $\kappa_a^b$.
If $\mu_{Zh}^0$ is found to be noticeably less than 1, the degeneracy of the $Zb\bar{b}$ couplings found interpreting the precision electroweak data at LEP and SLC can be broken and 
new physics beyond the SM must exist.

\vspace{3mm}
\noindent{\bf Acknowledgements.}
This work is partially supported by the U.S. Department of Energy, Office of Science, Office of Nuclear Physics, under Contract DE-AC52-06NA25396, [under an Early Career Research Award (C. Lee),] and through the LANL/LDRD Program, as well as the U.S.~National Science Foundation
under Grant No.~PHY-2013791. C.-P.~Yuan is also grateful for the support from the Wu-Ki Tung endowed chair in particle physics.

\bibliographystyle{apsrev}
\bibliography{reference}

\end{document}